\documentclass[final,1p,times]{elsarticle}
\usepackage{setspace}
\usepackage{graphics,amssymb,amsfonts, multirow,booktabs,amsmath,hyperref,epsfig,color,eurosym, amstext,color, dcolumn,adjustbox,lscape, placeins, graphicx}
\graphicspath{{figure/}}
\journal{arXiv}

\usepackage{makecell}
\usepackage[singlelinecheck=false]{caption}
\DeclareCaptionFont{tenpt}{\fontsize{10pt}{11pt}\selectfont #1}
\usepackage{float}
\hypersetup{pdfauthor={Name}}
\usepackage{subcaption}
\usepackage{endnotes}
\biboptions{authoryear}
\setcitestyle{aysep={,},yysep={;}}
\usepackage{pdflscape}
\usepackage{placeins}
\usepackage{graphicx}

\begin{document}
\begin{frontmatter}
\title{Redefining Urban Centrality: Integrating Economic Complexity Indices into Central Place Theory}

\author[a]{Jonghyun Kim}
\author[a,b,c]{Donghyeon Yu}
\author[a]{Hyoji Choi} 
\author[a,b]{Dongwoo Seo}
\author[a,c,d]{Bogang Jun \corref{cor1}}

\cortext[cor1]{\emph{Corresponding}: bogang.jun@inha.ac.kr}

\address[a]{Research Center for Small Businesses Ecosystem, Inha University, Incheon 22212, South Korea}
\address[b]{Department of Statistics, Inha University, Incheon 22212, South Korea}
\address[c]{Department of Data Science, Inha University, Incheon 22212, South Korea}
\address[d]{Department of Economics, Inha University, Incheon 22212, South Korea}

\begin{abstract} 
This study introduces a metric designed to measure urban structures through the economic complexity lens, building on the foundational theories of urban spatial structure, the Central Place Theory (CPT) (Christaller, 1933). Despite the significant contribution in the field of urban studies and geography, CPT has limited in suggesting an index that captures its key ideas. By analyzing various urban big data of Seoul, we demonstrate that PCI and ECI effectively identify the key ideas of CPT, capturing the spatial structure of a city that associated with the distribution of economic activities, infrastructure, and market orientation in line with the CPT. These metrics for urban centrality offer a modern approach to understanding the Central Place Theory and tool for urban planning and regional economic strategies without privacy issues.
\end{abstract}

\end{frontmatter}

\section{Introduction}

Understanding the dynamic nature of urban structures is crucial for elucidating the economies of a city and for implementing effective urban planning policies. The urban areas have transformed their structure ceaselessly. The structure, for example, co-evolved with the introduction of new transportation and logistics systems, such as railways in the 19th century and automobiles in the early 20th century, along with the emergence of new industries~\citep{anas1998urban}. These transformations underscore the critical need to comprehend the underlying spatial structures, as these structures are associated with the distribution of economic activities, infrastructure, and market orientations—key components in urban planning and regional economic strategies. To keep pace with these evolutionary changes, it is imperative to develop metrics that can accurately measure and reflect the current status of a region's urban structure.

This study introduces a metric designed to measure urban structures through the economic complexity lens, building on the foundational theories of urban spatial structure, particularly Central Place Theory (CPT) ~\citep{christaller1933zentralen}. Proposed by Walter Christaller in the 1930s, CPT has significantly influenced the field of urban studies and geography by offering a novel perspective on how central functions are distributed across an urban landscape, driven by consumer behavior and market forces. According to Christaller, settlements are organized in a hierarchical manner where central places fulfill specific roles by providing goods and services, which in turn shape the geographical and economic context, and, recursively, are shaped by the context. While Christaller’s CPT is well-known for its hexagonal market boundaries, his primary focus was, in fact, on the interaction between product centrality and location assuming a homogeneous space and uniformly distributed consumers~\citep{pacione2009urban}. Over the decades, CPT has provided an analytical framework for understanding the clustering of economic activities, particularly in retail trade and service industries, and has profoundly influenced urban and regional planning~\citep{mulligan2012central, bunge1962theoretical, boussauw2014short, dale2007changing, neal2011central, shearmur2015central}.


The incorporation of big data and advanced analytical techniques in recent years has reinvigorated CPT, allowing researchers to explore the theory’s micro-foundations through detailed data analysis, thus reaffirming its relevance in contemporary urban studies. \cite{van2018christaller}, for example, show the relationship between threshold and range of goods and the various central functions for the case of Louisville, Kentucky, by using Twitter data, confirming the micro-foundation of CPT. \cite{kii2023estimating} estimate the urban spatial structure based on remote sensing data of Tokyo, Japan, by estimating trip attraction. These researches have opened pathways to reconsider the main idea of CPT in the new light of the emergence of big data. Although the literature shows the applicability of CPT with big data, it is limited in developing a universal metric for CPT that highlights on a region and a product in terms of centrality. Moreover, literature tends to focus on the hexagonal configuration itself, rather than the interaction between the products and the locations that was a key emphasis of Christaller. 

\begin{figure}[!t]
\centering
\includegraphics[width=1.0\linewidth]{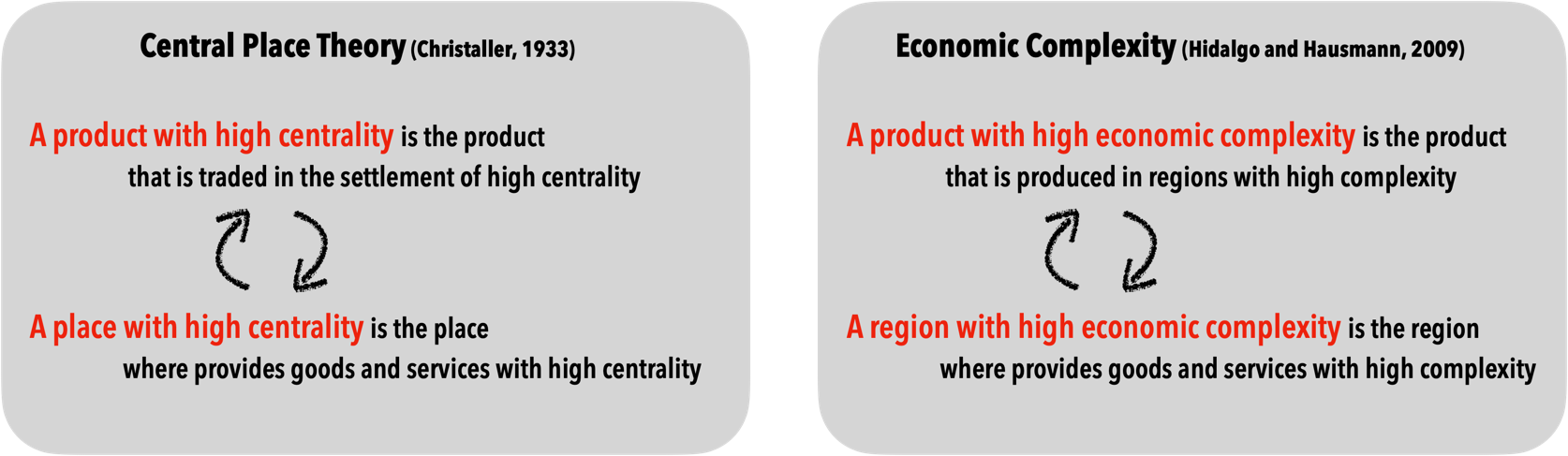}
\caption{Similarity of definitions between the Central Place theory~\citep{christaller1933zentralen} and the Economic Complexity~\citep{hidalgo2009building}}
\label{fig:concept}
\end{figure}

To suggest a metric for the universal measure of the centrality of products and locations, this study revisits the core principles of CPT, with more paying attention to the interaction between the the characteristics of goods and services, and the location that provides the goods and services. When we focus on this fundamental aspect of CPT, which is the recursive interaction between the centrality of products and places, we can find the resemblance of CPT to the concept of economic complexity as depicted in Figure~\ref{fig:concept}. 

The concept of economic complexity was introduced by \cite{hidalgo2009building}. In their seminal work, they derived two metrics, which are the Product Complexity Index(PCI) and the Economic Complexity Index(ECI), by looking at world trade data. The two complexity metrics show the recursive structure in their definition: a country with a high complexity measured by ECI is the country that produces complex products, while a product with a high complexity measured by PCI is the product that is produced in a country with high complexity. \cite{hidalgo2009building} derived the PCI and ECI by solving the eigenvalue problem of the recursive equations. Thanks to the flexibility and applicability of two metrics, ECI and PCI, it has become popular tools in economic geography, international development, and innovation studies~\citep{hidalgo2021economic}. 

In this work, we suggest a more universal way of measuring the centrality of products and locations in a city by combining the theoretical framework by Christaller and recent advancements in the field of Economic Complexity~\citep{hidalgo2009building, hidalgo2021economic}. We aim to show how the ECI and PCI reveal the urban structure in Seoul, which is a megacity, combining various big data, such as data on all the geo-locations of small businesses in Seoul; residential, labor, and floating population data; and credit card data with high spatial resolutions. Specifically, our new measures for the centrality of products and places, PCI of a product and ECI of a region, are calculated by using data on the geo-location of small businesses, which can be free from privacy issues so that it can be applied for the other cities easily. 


\section{Results}
\subsection*{\underline{Detecting the spatial unit of analysis}}
\label{section:spatial unit}
To explore the ECI and PCI of an urban area, we look at Seoul, which is a megacity, exhibiting a gross regional domestic product (GRDP) of $\$3,739$ million (20\% of all GDP) and involving 2,860,562 households, 823,624 firms, 5,226,997 workers and 9,662,041 residents in 2019. Then, based on the location data of the small businesses shops, we define the spatial unit of analysis. While administrative districts are often used as a spatial unit of analysis, this administrative boundary within the city level does not determine consumption behavior. Therefore, we define the spatial unit of analysis based on data showing how the small businesses are located~\citep{hidalgo2020amenity, jun2022economic}, resulting in 523 amenity clusters, whose average radius is 241 meters. The following analysis is based on this detected spatial unit of analysis that is depicted with polygons in Figure~\ref{fig:eci_sm_label}.

\subsection*{\underline{The optimal distance among markets over products}}
We investigate the relationship between PCI and the spatial range of the market, using PCI in two ways. One focuses on products and services and examines their market boundaries--which reflect Christaller's concept of the demand \textit{threshold}, the minimum number of customers necessary for a central function to exist--by using the location data of all the small businesses in Seoul. The demand threshold is captured by the minimum distance among markets of a product. The other approach focuses on consumer behavior by using credit card data, by looking at how the distance between consumers' residential area and their places of their consumption differs over products. This reflects the maximum distance that a consumer is willing to travel to obtain a central function (the \textit{range} in Christaller's concept). We construct two empirical models, which are Equation~\eqref{eq:dist_sm} and \eqref{eq:dist_bc}, to investigate economic-geographical characteristics of central place goods and central places in terms of the PCI and ECI. 


First, the demand threshold in Christaller's idea is captured by the minimum distance among markets of a product. To calculate the minimum distance among markets, we start with calculating Revealed Comparative Advantage of product $p$ in cluster $i$ $(RCA = \frac{shops_{cp}/\sum_{p}shops_{cp}}{\sum_{c}shops_{cp}/\sum_{cp}shops_{cp}})$ by looking at the number of shops in the cluster, $shops_{cp}$, and figure out whether a product in a cluster has the comparative advantage~\citep{Balassa1965}. $RCA_{cp}$ is greater or equal to 1 when the share of production of cluster on a given product $p$ is larger than the share of that product on the entire city. When $RCA_{cp}$ is greater or equal to 1, we regard the market of product $p$ exists in cluster $c$. Then, we look at the distance between every pairs of the market and pick the minimum distance among them. That is the minimum distance between clusters for a product $p$, $Dist_{p,c,c'}$, which is our dependent variable. We construct the empirical specification in Equation~\eqref{eq:dist_sm} to investigate the spatial range of market by each product:
\begin{align}
    \label{eq:dist_sm}
    \begin{split}
    Dist_{p,c, c'} &= \beta_{0} + \beta_1 PCI_{p} + \beta_2 \Delta{ECI_{c,c'}} + \beta_3 \Delta{Diversity_{c,c'}} + B \cdot X  \\
    & + \mu_{location} + \mu_{industry} + \varepsilon_{p,c, c'}
    \end{split}
\end{align} where the dependent variable, $Dist_{p,c.c'}$, is the minimum distance between two spatial units that shows comparative advantages in product $p$. Our variable of interests is the product complexity, $PCI_{p}$. We also control for the difference in economic complexity of two spatial units, $\Delta{ECI_{c, c'}}$, and product diversity, $\Delta{Diversity_{c,c'}}$. $B \cdot X$ includes other control variables, such as the demographic characteristics, difference in labor, floating, and residential population in thousands of people, and the difference in land price between the spatial units. Additionally, we add vectors of dummy variables to control location-specific and industry-specific fixed effects and $\varepsilon_{p,c, c'}$ is the error term~\footnote{The administrative district dummy variable was measured at the ward level \texttt{gu}, which is an intermediate step in the city's administrative division. The dummy variable for industry classification is measured at the highest level, with a total of nine classifications.}. The result is depicted in Columns (1) and (2) of Table~\ref{reg:mainresult}. 

\begin{figure*}[!t]
    \centering
    \includegraphics[width=1.0\linewidth]{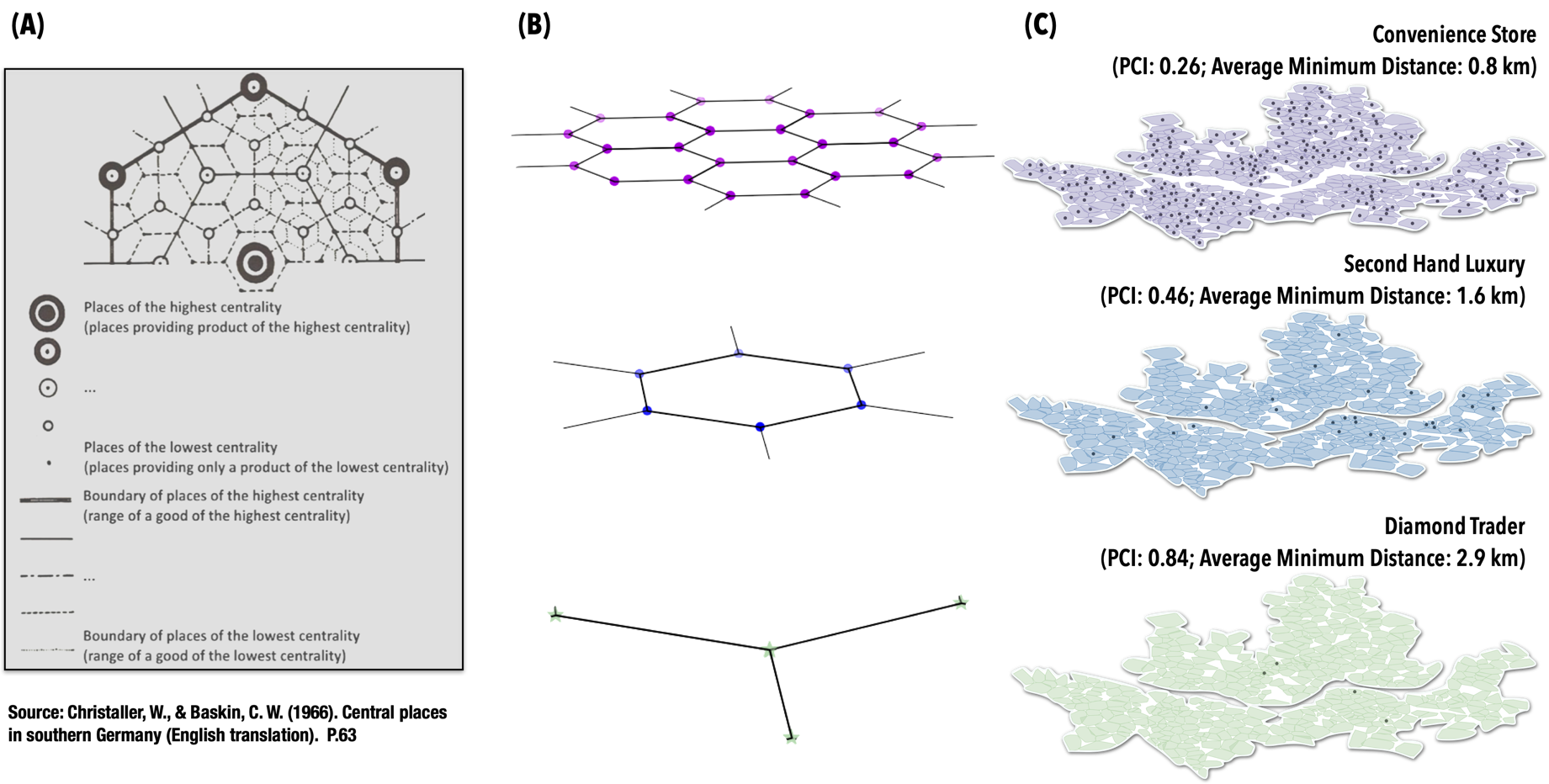}
    \caption{Market boundaries over the level of PCI. (A) Hexagonal configuration the product centrality and (B) schematic market boundaries over different products suggested by Christaller (1933). (C) Calculated market boundaries of convenience store (PCI = 0.26, average market distance = 0.8 $km$), Second handed luxury goods (PCI = 0.46, average market distance = 1.6 $km$), and Diamond trader shop (PCI = 0.84, average market distance = 2.9 $km$). Black dots represent the centers of small business clusters that exhibit the comparative advantage of the product.}
    \label{fig:framework}
\end{figure*}

Before moving to the second empirical specification, we visualize our result on the relationship between the market boundaries of products and their PCI as shown in Figure~\ref{fig:framework}. Figure~\ref{fig:framework} (A) is the conceptual framework suggested by ~\cite{christaller1933zentralen}. He argues that the spatial configuration of a city depends on the centrality of places. In turn, central places are those providing products and services with high centrality according to the iterative structure of his argument. Based on Christaller's idea, we can draw the schematic market boundary of different products as depicted in Figure~\ref{fig:framework} (B). This shows that the more central products are likely to have a longer distance between the markets. As expected by the conceptual framework of \cite{christaller1933zentralen}, Figure~\ref{fig:framework} (C) shows that the more complex product, which corresponds with the higher centrality, tends to exhibit the larger market boundary. For example, in the case of convenience stores, which have lower centrality, the average distance among the market is 0.8 $km$, and its product complexity is 0.26. In contrast, diamond trader shops, which have higher centrality, have a long market distance, which is 2.9 $km$, with a PCI of 0.84. Although the spatial configuration is not the exact hexagonal shape as the conceptual framework expected in Figure~\ref{fig:framework} (A) and (B), which is based on the assumption with homogeneous space and uniformly distributed consumer, our results show the materialized world of the idea. 

Next, we construct the second empirical specification, Equation~\eqref{eq:dist_bc}, to explore relationships between consumer's purchasing mobility and the product centrality measured by PCI. Because of privacy issue, tracking individual consumer's purchasing behavior is impossible using our data. Instead, given information aggregated by purchasing and residential location, and the demographic information on consumers, we constructed consumer group $i$ by looking at their age (aggregated by 10s), gender, purchasing and residential location based on 50m $\times$ 50m cell units. Our empirical specification to see the purchasing behavior of consumer group $i$ for product $p$ is following: 
\begin{align}
    \label{eq:dist_bc}
    \begin{split}
        Dist_{i,d} &= \beta_{0} + \beta_{1} PCI_{p} + \beta_{2} Count_{i,d} + \beta_{3} Female_{i} + B \cdot X\\
        &+ \mu_{location} + \mu_{industry} + \varepsilon_{i,p}
    \end{split}
\end{align} where the dependent variable, $Dist_{i,p}$, is the travel distance of consumers group $i$, who purchase goods of product type $p$ in a region from their residential area. Our main variable of interest is again product complexity $PCI_{p}$, while $Count_{i,p}$ is the aggregated purchase count for product $p$ by consumer group $i$. $Female_{i}$ is a binary indicator of consumers' gender that takes 1 if consumers are female and 0 otherwise. $B \cdot X$ includes demographic controls, such as age groups of consumers. We also add the location and industry fixed effects, which are $\mu_{location}$ and $\mu_{industry}$, to control for the time-invariant characteristics of consumers' residential location and the industry field of the goods\footnote{The administrative district dummy variable was measured at the ward level \texttt{gu}, which is an intermediate step in the city's administrative division. The dummy variable for industry classification is measured at the highest level, with a total of nine classifications.}. $\varepsilon_{p,o}$ is the error term. The result is depicted in Column (3) and (4) of Table~\ref{reg:mainresult}. 

As reported in Table~\ref{reg:mainresult}, a high PCI product is likely to have a broader market boundary, pulling consumers from a further distance. In Columns (1) and (2), the PCI is significantly and positively correlated with the minimum distance between two markets of product $p$. Since the ECI and PCI are normalized, the coefficients of $PCI_p$ in Columns (1) and (2) indicate that the minimum distance between markets is, on average, 0.297- 0.324 km further apart when the PCI value increases 1 unit of standard deviation. The difference between the case of minimum and maximum PCI is 2.1-2.3 kilometers further apart.

\begin{table}[h] \centering 
\caption{Relationship between the market boundary and the PCI of products} 
\label{reg:mainresult} 
\scalebox{0.75}{
    \begin{tabular}{@{\extracolsep{5pt}}lcccc} 
        \\[-1.8ex]\hline 
        \hline \\[-1.8ex] 
        & \multicolumn{2}{c}{\textit{Market boundary focusing on product}} & \multicolumn{2}{c}{\textit{Market boundary focusing on consumer's behavior}}\\ 
        \cline{2-3} \cline{4-5}
        \\[-1.8ex] & (1) & (2) & (3) & (4) \\ 
        \hline \\[-1.8ex]  
        $PCI_{p}$ & 2.272$^{***}$ & 2.082$^{**}$ & 8.942$^{***}$ & 9.228$^{***}$\\ 
          & (0.092) & (0.089)  & (0.011) & (0.011) \\ 
          & & & & \\
        $\Delta{ECI_{c,c'}}$ &  & 0.583$^{***}$ & & \\ 
        & & (0.069) & & \\
        & & & & \\
        $\Delta{D_{c,c'}}$ &  & 0.001$^{***}$ & & \\ 
        &  & (0.0005) & & \\ 
        & & & & \\
        $Count_{i,d}$ & & & & $-$0.388$^{***}$\\ 
          & & & & (0.001) \\ 
          & & & & \\
        $Female_{i}$ & & & 0.195$^{***}$ & 0.098$^{***}$ \\ 
          & & & (0.004) & (0.004) \\ 
          & & & & \\
        \textit{Demographics controls} & \checkmark  & \checkmark & \checkmark & \checkmark \\  
           & & & & \\ 
        \textit{Land price controls} & \checkmark & \checkmark & & \\  
           & & & & \\    
        $FE_{industry}$ & \checkmark & \checkmark  & \checkmark & \checkmark \\ 
        & & & & \\ 
        $FE_{location}$ & \checkmark & \checkmark  & \checkmark & \checkmark \\ 
        & & & & \\    
        $FE_{consumer age}$ & & & \checkmark & \checkmark \\ 
        & & & & \\  
        \hline \\[-1.8ex] 
        Observations & 22,215 & 22,215 & 4,201,653 & 4,201,653 \\ 
        R$^{2}$ & 0.220 & 0.274 & 0.277 & 0.294 \\ 
        Adjusted R$^{2}$ & 0.201 & 0.256 & 0.277 & 0.294 \\ 
        Residual Std. Error & 0.9156 & 0.883  & 4.200 & 4.155 \\ 
        F Statistic & 11.169$^{***}$ & 14.768$^{***}$ & 39,316.160$^{***}$ & 41,569.570$^{***}$ \\ 
        \hline 
        \hline \\[-1.8ex] 
        \textit{Note:}  & \multicolumn{2}{r}{$^{*}$p$<$0.1; $^{**}$p$<$0.05; $^{***}$p$<$0.01} \\ 
    \end{tabular}
}
\end{table} 

Column (3) and (4) of Table~\ref{reg:mainresult} shows the second results that are associated with the empirical specification of Equation~\ref{eq:dist_bc}. Again, the dependent variable is the distance between the consumers' residential area and their consumption places. As shown in Columns (3) and (4), a product with a higher PCI is likely to be purchased by consumers who are visiting the market from longer apart. Interestingly, as shown by the positive and significant coefficient of $Female_{i}$, the mobility of female consumers is higher than that of male consumers. The results in Columns (3) and (4) of Table \ref{reg:mainresult} show that on average, consumers travel 8.9 -- 9.2 $km$ further to purchase the more complex product (or service), when the PCI of the product increases 1 unit of standard deviation, and as suggested \cite{christaller1933zentralen}, the higher centrality goods measured by the PCI are associated with the longer boundary of the market of the product. 


\subsection*{\underline{Economic complexity of a region reveals the central place of a city}}
\begin{figure*}[!h]
    \centering
    \includegraphics[width=1.0\linewidth]{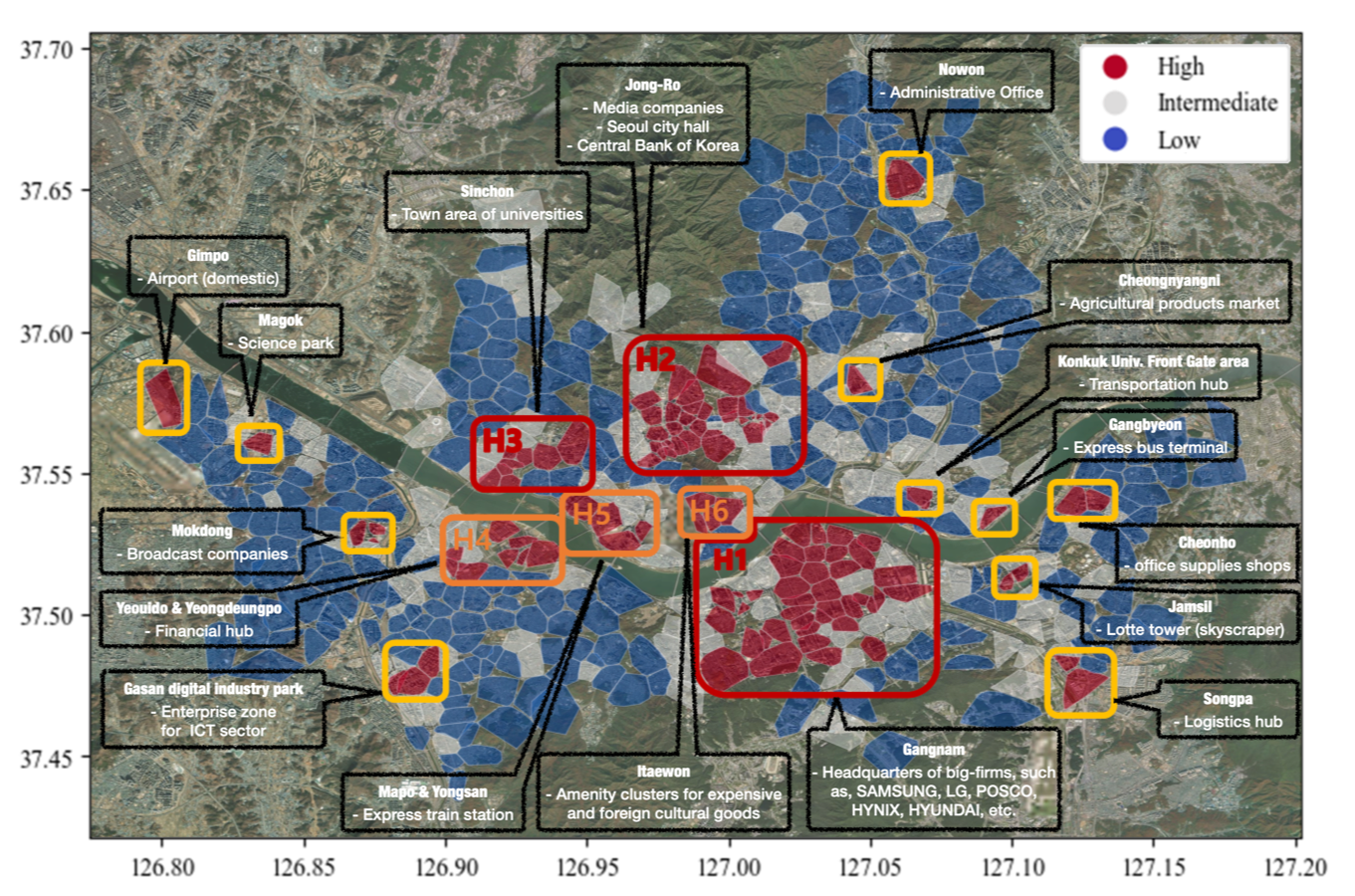}
    \caption{ Economic complexity of amenity clusters in Seoul. Amenity clusters with high, medium, and low level of ECI are in red, gray, and blue, respectively. }
    \label{fig:eci_sm_label}
\end{figure*}

We visualize ECI for each spatial unit in Seoul in Figure~\ref{fig:eci_sm_label}. Although we calculate the ECI by just looking at the locations of small businesses' shops in a city, interestingly, the result shows that high ECI regions, which are in red, are mostly well-known place for city's economically and politically central functions. For example, spatial units labelled $H1$ in Figure~\ref{fig:eci_sm_label}, primarily include the headquarters of major conglomerates or big companies. In these area, which has high ECI, complex products and services are more likely to provided/traded, as suggested by \cite{christaller1933zentralen}. Likewise, other clusters with high ECI, which are in red, are also responsible for central function of city Seoul: regions labelled H2 include Seoul city hall and leading media companies, those with H3 are university town areas, those with H4 are financial hub in nation, those with H5 are the small businesses' cluster of small electronic shops in \textit{Yongsan}, and those with H6 are the area where are the hub of foreign cultures or expensive amenity goods in \textit{Iteawon}. 
\begin{figure*}
    \centering
    \includegraphics[width=0.9\linewidth]{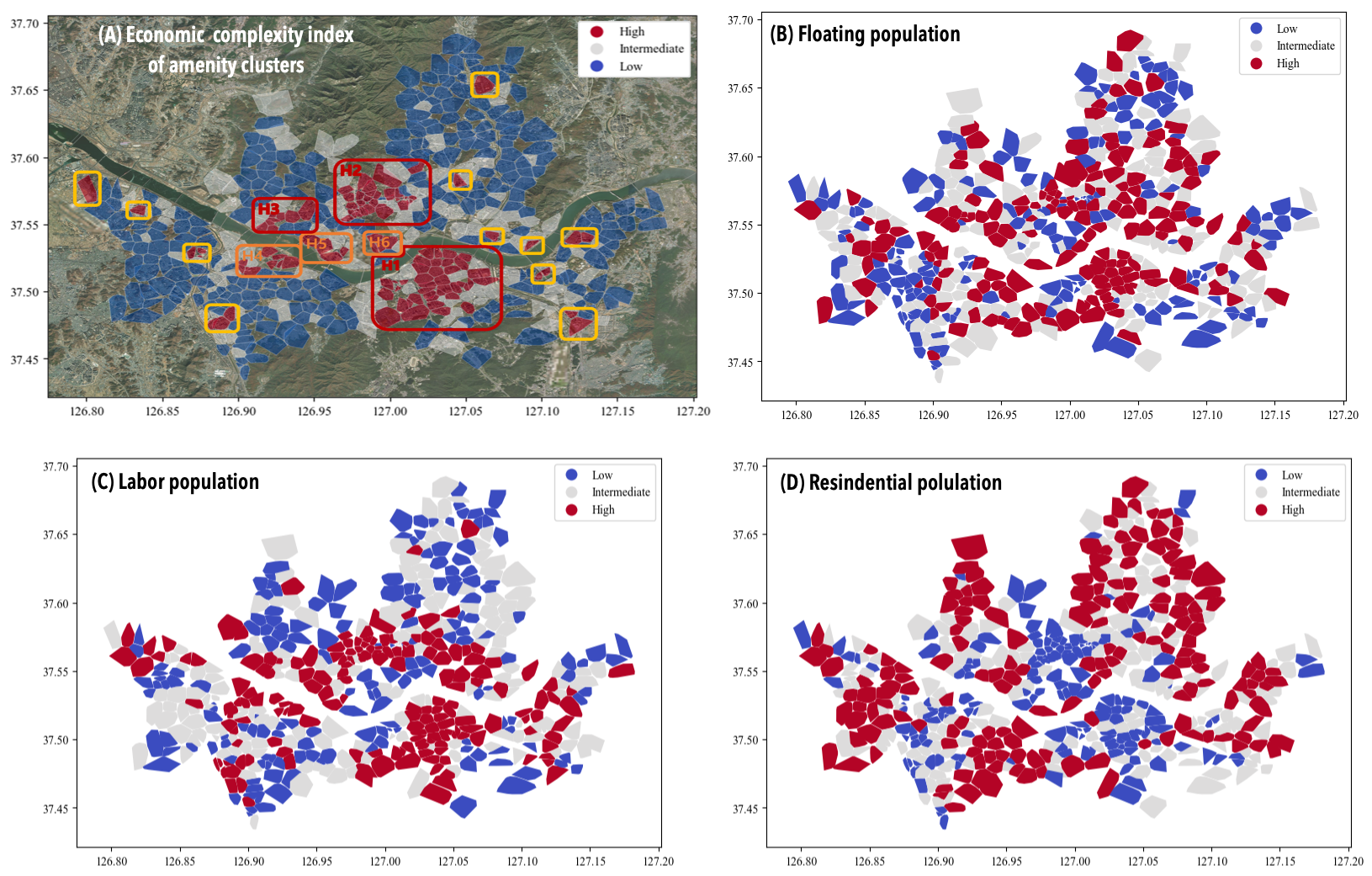}
    \caption{The level of ECI (A) and the density of floating (B), labor (C), and residential (D) population in Seoul. The high, medium, and low level of the values in a cluster are depicted in red, gray, and blue color, respectively.}
    \label{fig:popdensity}
\end{figure*}
Other regions with high ECI with yellow boxes are located in relatively peripheral area in Seoul but they are also serve for central city function with including central facilities, such as domestic airport, science park, cluster of private academies, enterprise zone for ICT companies, central market of agricultural and marine product in a country, administrative office of sub-district of Seoul, a central express bus terminal, amusement park, market for office supplies, and logistics $\&$ distribution hub. As demonstrated previously, the spatial concentration of economically and functionally important facilities on high ECI regions. 

Further, focusing of production side, we explore another spatial concentration on the high ECI regions, since according to \cite{christaller1933zentralen}, not only do central places play a crucial role in consumption, but they are also pivotal in production. To examine this assertion, we analyzed the spatial concentration patterns of the labor force and potential consumers in Seoul by visualizing the number of labor, floating, and residential populations within the city's clusters. These patterns are depicted in Figure~\ref{fig:popdensity}. The results show that regions with high ECI correspond to areas with bigger number of both floating and labor populations. However, these high ECI regions do not necessarily coincide with a bigger number of residential population. This finding supports Christaller's theory by illustrating that central places primarily support production activities, whereas surrounding areas with lower centrality predominantly serve residential functions.

\begin{table}[!htbp] \centering 
  \caption{Correlations between the various characteristics of an area and the ECI of a region. $ECI_c$ is the economic complexity of the cluster, $Diversity_c$ is the diversity of small business shops, $Shops_c$ is the number of shops in the cluster, and $Price_c$ is the average land price of the area. $Labor_c$, $Float_c$, and $Resi_c$ are labor, floating, and residential population numbers, respectively.} 
  \resizebox{\textwidth}{!}{
  \label{table:corr_pop} 
\begin{tabular}{@{\extracolsep{5pt}} lcccccccc} 
\\[-1.8ex]\hline 
\hline \\[-1.8ex] 
 & $ECI_{c}$ & $Diversity_{c}$ & $Shops_{c}$  & $Labor_{c}$ & $Float_{c}$ & $Resi_{c}$ & $Price_{c}$  \\
\hline \\[-1.8ex] 
 $ECI_{c}$ & $1$ &  &  &  &  &  \\ 
 $Diversity_{c}$ & 0.056& $1$ &  &  &  & \\ 
 $Shops_{c}$ & 0.449 & 0.706 & $1$ &  &  &  &   \\ 
 $Labor_{c}$ & 0.564 & 0.250 & 0.384 & $1$ &  &  &  \\ 
 $Float_{c}$ & 0.267 & 0.638 & 0.667 & 0.320 & $1$  &  &   \\ 
 $Resi_{c}$ & $-$0.574 & 0.413 & 0.184 & $-$0.281 & 0.271 & $1$ &   \\ 
 $Price_{c}$ & 0.661 & $-$0.034 & 0.222 & 0.442 & 0.064 & $-$0.481 & $1$  \\ 
\hline \\[-1.8ex] 
\end{tabular} %
}
\end{table} 

To delve into the production side, we check the Pearson correlation coefficients between the population type in regions and the ECI of them (Table~\ref{table:corr_pop}), and the point-biserial correlation between the labor share of industrial sector in a region and the ECI in the region\footnote{The industrial sectors are classified by the Korean Standard Industrial Classification.} (Table~\ref{table:corr_laborshare}). Table~\ref{table:corr_pop} shows that the high correlation between the ECI and the labor population of the area, while negative correlation between the ECI and the residential population. Furthermore, the place with high ECI is corresponds with relatively higher land price as well. Interestingly, the diversity of amenity shops is not necessarily correspond with the high ECI of the area. 

\begin{table}[!h] 
\centering 
  \caption{Correlation coefficients between labor share of industrial sector and the ECI of a region} 
  \label{table:corr_laborshare}
\resizebox{\textwidth}{!}{
\begin{tabular}{lrrr}
\toprule
 & \multicolumn{3}{c}{The level of ECI in a region} \\ 
Industrial sector & High & Intermediate & Low \\
\midrule
Financial and insurance & 0.386 & 0.036 & -0.355 \\
Professional, scientific and technical services & 0.383 & 0.138 & -0.440 \\
Information and communication & 0.280 & 0.208 & -0.415 \\
Wholesale and retail trade & 0.222 & 0.012 & -0.196 \\
Business facilities management and business support services; rental and leasing activities & 0.075 & 0.180 & -0.219 \\
Electricity, gas, steam and air conditioning supply & 0.042 & -0.056 & 0.013 \\
Accommodation and food service activities & 0.035 & -0.083 & 0.042 \\
Agriculture, forestry and fishing & 0.014 & -0.024 & 0.009 \\
Public administration and defence; compulsory social security & -0.014 & -0.026 & 0.034 \\
Mining and quarrying & -0.028 & 0.084 & -0.049 \\
Transportation and storage & -0.099 & -0.135 & 0.199 \\
Construction & -0.109 & 0.077 & 0.025 \\
Real estate activities & -0.131 & -0.050 & 0.153 \\
Manufacturing & -0.162 & -0.060 & 0.188 \\
Water supply; sewage, waste management, materials recovery & -0.193 & -0.065 & 0.218 \\
Arts, sports and recreation related services & -0.224 & -0.000 & 0.187 \\
Education & -0.311 & -0.059 & 0.3152\\
Human health and social work activities & -0.378 & -0.193 & 0.483 \\
Membership organizations, repair and other personal services  & -0.386 & -0.167 & 0.468 \\
\bottomrule
\end{tabular}
}
\end{table} 
Table~\ref{table:corr_laborshare} shows that population working in ``financial and insurance sector'' and ``professional, scientific and technical services'' are more likely to be concentrated in high ECI regions, by showing relatively high value of correlation 0.386 and 0.383, respectively. By contrast, labor shares in construction, manufacturing, and sectors related with infra structure, which are a bit far from the central role of a city, are higher in regions with relatively low ECI, with scoring negative correlation. 

According to Table~\ref{table:corr_pop} and \ref{table:corr_laborshare}, the high ECI regions exhibit the high ECI not only in terms of city function as shown in Figure~\ref{fig:eci_sm_label} but also in production side. Our results confirms the hypothesis suggested by Christaller that the cenral places plays a crucial role in consumption and at the same time, in production as well.

\section{Discussion and Conclusion}
In the preface of his doctoral thesis, Christaller stated that the purpose of Central Place Theory (CPT) is to analyze the economic and geographical principles that determine the spatial arrangement of urban functions. Rather than merely describing these arrangements, as typified by the hexagonal configuration, he emphasized the interactions between central functions—represented by sets of goods and services—and their centrality within the city. Despite of the theoretical contribution of CPT, it has been limited to apply for the real world, since CPT was developed with assumption the uniformly distributed consumer in a homogeneous space. When focusing on the key principle of CPT, we could find the similarity between the concept of economic complexity, which is already widely used thanks to its flexibility and applicability, and suggest the two economic complexity indices, PCI and ECI, as a metric for the centrality in CPT. We calculate the PCI of products and the ECI of places by using the geo-location data of small businesses in Seoul and test whether these two metrics well capture the centrality of products and places. 

First, to capture the effect of PCI of a product on the market boundary of the product, we calculate the minimum distance among markets of the product by using the shop location data, and see the effect of the PCI on the minimum market distance. Our result show that a high PCI product is likely to have a broader market boundary. When 1 unit of the standard deviation of PCI increases, the minimum distance among the product's markets increases, on average, 0.297-0.324 km. The difference in market distance between a product with the minimum PCI and that with the maximum PCI is 2.1-2.3 km. In addition, to capture the demand threshold of a product, which is the minimum number of customers necessary for a central function to exist, we look at the effect of PCI on the distance between the residential area and the shop the consumption behavior happen by using the card transaction data. Our results show that a consumer who purchase a high PCI product is more likely to be from further away from the shop. When the one standard deviation of the PCI of products increases, consumers travel 1.782-1.838 km further to purchase the product. The difference in market range between a product with the minimum PCI and that with the maximum PCI is 8.2--9.2 km. 

Second, following the logical process in the Christaller's theory, we look at the economic complexity of places. According to him, the centrality of location is based on the accumulated functions provided by products and services on the place, given that there exist various products in the place. We expected that the recursive structure of economic complexity index well capture this aspect and see whether the ECI of places successfully identify the central places in terms of city function and the production. When we visualized the ECI of places, we found that high ECI places are the well-known for Seoul's economically and politically central places including, for example, the central bank of Korea, a terminal for high speed rail and express bus terminal, financial hub, headquarters of big firm, like Samsung. When checking the production side by looking at the population composition of a place and the industrial sector that the labor share is working at, we find that the high ECI places play a central role for more complex economic activities. Furthermore, the high ECI places are strongly correlated with the labor population, especially those associated with knowledge intensive industries. Our results confirm that ECI is well representing the level of centrality of places. 

\begin{figure*} [h] \centering 
    \centering
      \includegraphics[width=1\linewidth]{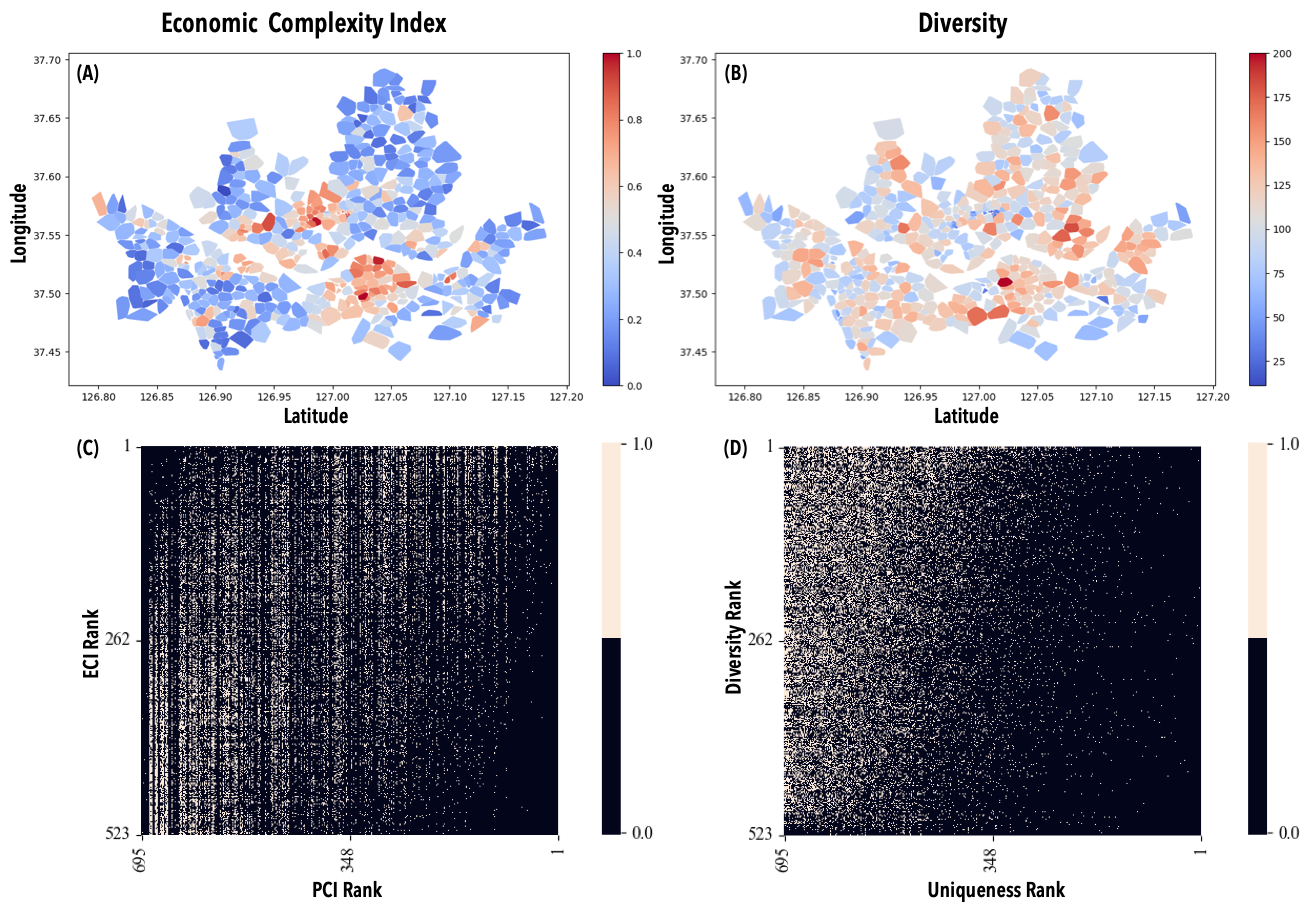}
    \caption{Robustness of economic complexity indices as a measure for the centrality using the location data of all shops in Seoul. (A) ECI and (B) Diversity over the spatial units in Seoul. (C) Contingency matrix between ECI rank and PCI rank and (D) between Diversity rank and Uniqueness rank. }
    \label{fig:robustness}
\end{figure*}

One might argue that an alternative metric, such as the diversity of economic activity in a place, could effectively represent the centrality of a place. This notion could be supported by Christaller's argument that the most unique goods are produced in the most central locations, whereas ubiquitous goods are produced universally, independent of a place's centrality. To explore this, we examined the diversity of goods and their spatial distribution in Seoul, comparing to these findings with the results of PCI and ECI. As illustrated in Figures~\ref{fig:robustness} (A) and (B), the regional diversity rankings in Seoul present a different picture to that depicted by the complexity index, with regions of high diversity not necessarily correlating with the central functions of Seoul.

Further analysis involved examining a contingency matrix for the diversity versus the uniqueness of goods, as shown in Figure~\ref{fig:robustness} (D), to investigate the recursive relationship between uniqueness and diversity. The dark area on the left side of Figure~\ref{fig:robustness} (D), regardless of the level of diversity, indicates that the uniqueness of products does not correlate with diversity. A recursive structure, if present, would likely show products in the upper right corner of the matrix, indicating regions where high uniqueness products are produced alongside high diversity, as the relationship between PCI and ECI in Figure~\ref{fig:robustness} (C). For the principle of low PCI/uniqueness products being produced in low ECI/diversity regions to hold, the top left corner of the contingency matrix would need to be dark. However, this is observed only in the case of the economic complexity index (Figure~\ref{fig:robustness} (C)), but not for diversity. These findings confirm that the economic complexity indices, PCI and ECI, effectively embody the recursive structure of centrality defined in CPT.

Taken together, we present a metric to measure the centrality of a product and a region with borrowing the theoretical foundation of Christaller's CPT. Given the core principle of CPT is the interaction between the centrality of products and places, our metrics, stemming from the field of economic complexity, are conceptually consistent with the centrality of products and places. Our findings supports our arguments that PCI of products are positively correlated with the market size of the products, and high ECI places is the place that provides the central city functions and complex products. Considering that we use the location data of shops that are free from the data privacy issue, our findings can contribute to the body of literature on urban economy and policy makers by capturing the dynamic nature of urban structure more precisely and easily.

\section{Methods}
\subsection*{\underline{Detecting the spatial unit of analysis}}
By using the location data of all small business shops in Seoul, first, we calculate the effective number of small businesses, $ A_i $ and detect the amenity dense neighborhood~\cite{hidalgo2020amenity, jun2022economic}. 
\begin{equation}
 A_i = \sum^{N}_{j=1} e^{-\gamma d_{ij}}
\end{equation} where $N$ denotes the total number of stores in the city and $d_{ij}$ is the geodesic distance between stores $i$ and $j$. With the stand point of small business shop $i$, the effect of the existence of other store is diminishing over distance. $\gamma$, decay parameter capture this diminishing influence of distant stores. In our analysis, we set $\gamma$ into 7.58. 

Second, by looking the spatial distribution of the effective number of shops, we identify the local peaks of $A_i$ on the map. Then, as capturing the center of the cluster with the local peak, we allocate neigh by shops to the cluster until the boundaries of the cluster are overlapped, resulting in 523 amenity clusters, whose average radius is 241 meters. Following analysis is based on this detected spatial unit of analysis that are depicted in  Figure~\ref{fig:eci_sm_label}.

\subsection*{\underline{Data}}
On the detected spatial unit of analysis, we accumulated various big data. To detect the amenity clusters and examine the composition of various amenity types they provides, we use data on small businesses sourced from the Korean Small Enterprise and Market Service (KSEMS). This dataset encompasses detailed information on the location and business categories of 401,071 small businesses observed to be in operation during the period from March 2019 to June 2019. To investigate the urban structure without COVID-19 effects, we choose the data of 2019. 

To examine the structure of economic activities in urban areas, we use spatial statistics data on residential, labor, and floating populations in 2019 and credit card transactions in June 2019. First, the residential and labor population data is from \texttt{Korea National Statistical Office (KOSTAT)}. The residential and labor populations represent the number of people living in households and working within a given area based on 100m $\times$ 100m cell unit, respectively. Second, human mobility big data, a floating population, is collected from the cell phone activity of personal subscribers to \texttt{SK} Telecom, the biggest network provider in South Korea. Here, floating population refers to the daily average of people staying or moving within a given area based on 50m $\times$ 50m cell unit. 

Finally, we also use credit card data provided by \texttt{BC} Credit Card Corporation for the robustness check. The credit card data includes information on the total number and amount of purchases (in KRW), number of stores, location, and industry classification provided by \texttt{BC}. Due to privacy concerns, the information is provided with aggregated based on 50m $\times$ 50m cell unit. 

We reconfigure location information of the residential, labor, and floating population data and credit card data into our detected amenity clusters in Section~\ref{section:spatial unit}. Again, there are 523 small business clusters in our analysis.

\subsection*{\underline{Calculating complexity index of region and product}}

Motivated by the Christaller's conceptualization of place centrality~\citet{christaller1933zentralen}, whose bears resemblance to the principles underlying the Economic Complexity Index, we utilize economic complexity indices as proxies for the assessing the centrality of locations and goods. Figure~\ref{fig:concept} describe the concepts of centrality by Christaller and that of economic complexity, showing the similarity between the two in terms of recursive structure. \citet{hidalgo2009building} introduced new metric on economic complexity, with describing that complex economies are those involved in complex activities, while complex activities are those performed by complex economies~\citep{hidalgo2021economic}. By using the method of reflection, \citet{hidalgo2009building} derived the complexity indexes that the economic complexity index (ECI: the complexity of embedded production capability of a place), which is normalized average of the product complexity, and that the product complexity index (PCI: the complexity of required production capability for a product), which is normalized average of the economic complexity. Following \cite{hidalgo2009building}, we calculate ECI of a region and PCI of a product as shown in Equation~\ref{eq:sys_econ} and ~\ref{eq:sys_prod}. 

\begin{align}\label{eq:sys_econ}
\begin{split}
ECI_{c} &= K_c = \frac{1}{M_{c}} \sum_{p} M_{cp} K_{p}
\end{split}
\end{align}
\begin{align}\label{eq:sys_prod}
\begin{split}
PCI_{p} &= K_p = \frac{1}{M_{p}} \sum_{c} M_{cp} K_{c}
\end{split}
\end{align} where $M_{cp}$ is a binary matrix consisted of 0 or 1, representing subscript $c$ is cluster of small businesses, while $p$ is product. To determine an element of the matrix, we calculate Revealed Comparative Advantage of a product $p$ in the shops' cluster $c$ $(RCA = \frac{shops_{cp}/\sum_{p}shops_{cp}}{\sum_{c}shops_{cp}/\sum_{cp}shops_{cp}})$ by looking at the number of shops in a cluster, $shops_{cp}$~\citep{Balassa1965}. When $M_{cp}$ equals to 1, we regard a cluster $\texttt{c}$ has the comparative advantage in product $p$.\footnote{ We normalize the complexity indexes, between 0 and 1 for the convenience of interpreting results: $ECI_{c} = \frac{K_{c}-min(K_{c})}{max(K_{c})}$, $PCI_{p} = \frac{K_{p}-min(K_{p})}{max(K_{p})}$ }

\section*{Acknowledgement}
This work was supported by the National Research Foundation of Korea (NRF) Grant through the  Korea Government [Ministry of Science and ICT (MSIT)] under Grant NRF-2022R1A5A7033499. We also thank the support from Inha University.

\section*{Data Availability Statement}
The small business firm data is provided by \texttt{Korean Small Enterprise and Market Service (KSEMS)}. The labor and residential population data is provided by \texttt{Korea National Statistical Office (KOSTAT)}. The original small business firms and labor and residential population data can be found in the \texttt{OPENapi} portal: \url{https://www.data.go.kr/} and \url{https://sgis.kostat.go.kr/}. 

\linespread{1.5}
\newpage
\linespread{1.5}

\end{document}